\documentclass{PoS}
\newcommand{\beq}{\begin{eqnarray}}
\newcommand{\eeq}{\end{eqnarray}}

\newcommand{\ie}{{\it i.e.\ }}

\usepackage{epsfig}
\usepackage{graphicx}
\usepackage{latexsym}
\usepackage{wrapfig}

\title{Lattice QCD at finite density: \\ imaginary chemical potential}

\ShortTitle{Imaginary $\mu_B$}

\author{\speaker{Maria Paola Lombardo}\\
        Istituto Nazionale di Fisica Nucleare\\
        Italy\\
        E-mail: \email{lombardo@lnf.infn.it}}

\abstract{I describe the results for the critical line and the 
thermodynamics of different phases of QCD which have been obtained 
by  lattice simulations 
with an imaginary chemical potential. I review motivations
and merits of the different strategies -- Taylor expansion, 
Fourier analysis and Pade' approximants -- for analytic continuation
from  imaginary to real chemical potential.
I  consider  phenomenological models which can be easily
extended to the complex  chemical potential plane, thus 
affording a direct  comparison with lattice data at 
imaginary $\mu$: the hadronic phase and the high temperature limit
are amenable to a simple description, while a rather subtle
interplay between thermodynamics and critical behaviour emerges
in the hot phase close to $T_c$.} 

\FullConference{The 3rd edition of the International Workshop --- The Critical Point and Onset of Deconfinement --- \\
		 July 3-7 2006\\
		 Galileo Galilei Institute, Florence, Italy}

\begin{document}
\section{Introduction}

In principle, the lattice formulation
provides a rigorous framework for the study of the thermodynamics
of QCD. In practise, however, the lattice regularisation 
is usually combined with importance
sampling, which cannot be naively applied at nonzero baryon density,
where the quark determinant becomes complex~\cite{Kogut:1983ia}.

It has been realised  that this problem can be circumvented
in the high $T$, low $\mu$ part of the QCD phase diagram where 
one can take advantage of physical fluctuations. 
Interesting physical information can be obtained by computing the
derivatives with respect to $\mu$ at zero chemical potential
and high temperature 
\cite{Gottlieb:1988cq,Bernard:2002yd,Gavai:2003mf,Choe:2002mt,Gavai:2003nn,
Allton:2002zi,Allton:2003vx}.
Fodor and Katz proposed an improved reweighting affording a better
overlap between simulation and target ensembles, and a  
first estimate of the critical endpoint 
\cite{Fodor:2001au,Fodor:2001pe,Fodor:2002km,Fodor:2004nz}.
 In refs.~\cite{Dag,HasTou92,alford}  
the  imaginary chemical potential approach 
was advocated and exploited in connection with the 
canonical formalism. 
In ref.~\cite{mpl} it was proposed that the analytic continuation 
from imaginary chemical potential could be practical at 
high temperature, and the idea was tested  in the infinite coupling limit.
In refs.~\cite{hart1} the method was 
applied successfully to QCD in 2+1 dimensions.
In ref.~\cite{deForcrand:2002ci}
 it was proposed that the critical line itself can be analytically
continued and results for two flavor of staggered fermions were presented.
The scaling of the critical line and thermodynamics of the four fermion
models were studied in \cite{D'Elia:2002gd,deForcrand:2003hx,D'Elia:2004at}

The purpose of this note is to review early and new 
results obtained by use of the imaginary chemical potential 
approach: the basic idea, the applications, and the potentiality. 

The following Section is a short introduction into lattice QCD and the sign 
problem; Section 3  discusses the phase diagram in the $T, \mu^2$ plane; 
Section 4  offers a short summary of the canonical approach, while Section 5 
introduces the three series representations which have been used so far.
Next, a short collection  of results from models. The five central Sections  
are devoted  to the presentations and discussion of results: Sections 7 
focuses on the analytic continuation from lattice data to real chemical 
potential by use of a Taylor expansion, section 8 shows how to extend
the results towards larger $\mu$ / smaller temperatures by using
Pade' approximants. Sections 9 and 10 use phenomenological models 
extended to entire complex lane to parametrise thermodynamics in the 
different phases of QCD, section 11 shows how critical behaviour at
imaginary chemical potential can influence the thermodynamics of the
strong interactive Quark Gluon Plasma.

Introductory material and other surveys of results can be found
e.g. in refs. 
\cite{Muroya:2003qs,Lombardo:2004uy,Philipsen:2005mj,Schmidt:2006us}.

\section{The sign problem, and an imaginary $\mu_B$}

Let us remind ourselves
  how to introduce a chemical 
potential $\mu$
for a conserved charge {$\hat N$} in the density matrix $\hat \rho$
in the Grand Canonical 
formalism, which is
the one appropriate for a relativistic field theory:
\begin{eqnarray}
\hat \rho &=& e^{-(H - \mu \hat N)/T}   \\ 
\cal Z (T, \mu) & =& Tr \hat \rho = \int d \phi d \psi e^{-S(\phi, \psi)} 
\end{eqnarray}

The temperature $T$ on a lattice is the same as in 
the continuum: $T = 1 /N_ta$, $N_ta$ being the lattice extent
in the imaginary time direction (while, ideally, the lattice
spatial size should be infinite).
A lattice realisation 
of a finite density of baryons, instead, poses specific problems:
the naive discretization of the continuum expression 
$\mu \bar \psi \gamma_0 \psi$  
would  give an energy
$ \epsilon \propto \frac{\mu^2}{a^2}$ diverging 
in the continuum  (${a \to 0}$) limit.
The problem could be cured by introducing appropriate counterterms, however 
the analogy between $\mu$ and  an external field in the $0_{th}$ 
(temporal) direction offers a nicer solution by
considering the appropriate lattice conserved current \cite{Kogut:1983ia}. 
This amounts to the following
modification of the fermionic part of
the Lagrangian for the $0_{th}$direction  $L_F^0$:
\begin{equation}
 L_F^0(\mu) = \bar \psi_x \gamma_0 e^{\mu a}\psi_{x + \hat 0} -
      \bar \psi_{x + \hat 0} \gamma_0 e^{-\mu a}\psi_{x }
\end{equation}

We then integrate out fermions exactly, by taking advantage of the
bilinearity of the fermionic part of the Lagrangian
$L = L_{YM} + L_F = L_{YM} +  \bar \psi M (U) \psi$ :
\begin{equation} 
\int dU d \psi d \bar \psi {\cal Z} (T, \mu, \bar \psi, \psi, U)  = 
\int dU e ^{-(S_{YM}(U) - log(\det M))} 
\end{equation}

\begin{figure}
\centerline{
\includegraphics[width= 12truecm]{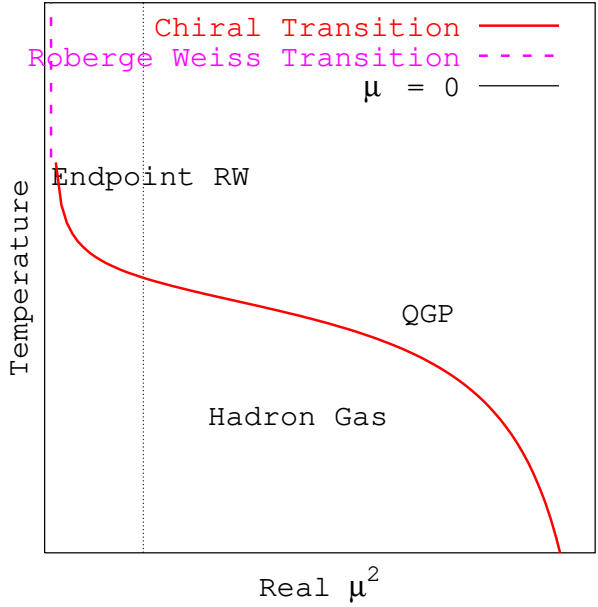} }
\caption{The phase diagram in the $T, \mu^2$ plane. Simulations are
possible in the $\mu^2 \le 0$ halfplane, and results in the physical
$\mu^2 \ge 0$ halfplane have to be inferred from the results. }
\label{fig:pd}
\end{figure}

When $\det M > 0$ the functional integral can be evaluated
with statistical methods, sampling the configurations according
to their importance $(S_{YM}(U) - log(\det M))$.
For this to be possible the would-be-measure  ($\det M $) has to be positive.

Consider now the relationship 
$M^\dagger (\mu_B) = - M (-\mu_B) $: this implies
that  reality is lost when $ {\rm Re} \mu \ne 0$.
Clearly, for real
$\mu$,  the imaginary part of the determinant cancels out in the statistical
ensemble, and it is even possible to cancel it exactly on
a finite number of configurations by considering the appropriate symmetry
transformation \cite{Alles:2002wh,Ambjorn:2002pz}; but still one has
to face a sign problem, and it is not clear in which dynamical region
this problem becomes significant \cite{Splittorff:2006vj}.

Consider instead  ${\rm Re} \mu = 0$ 
and ${\rm Im} \mu \ne 0$: in this case  
$M^\dagger (\mu_B) = - M (-\mu_B) $
implies $M^\dagger = - M^* $ and standard lattice simulations using
$M^{\dagger} M$ as a weight are possible. 

It has been shown by Roberge and Weiss~\cite{Roberge} that $Z(i\mu / T)$ 
is always periodic  $2 \pi /3$, for any physical temperature.
At  low $T$ strong coupling calculations predict a    smooth 
behaviour, whereas  at high $T$ weak coupling calculations predict 
discontinuities in the thermodynamics observables at 
$T = 2 \pi /3 (k + 1/2)$. 

This scenario for the phase diagram of QCD in 
the $T$ -- $i \mu_I$ plane  has been indeed confirmed by lattice studies
\cite{deForcrand:2002ci,D'Elia:2002gd} 
and model calculations \cite{Dumitru:2005ng}. It was also found that
the Roberge-Weiss line of discontinuities ends around a critical
temperature $T_{RW} > T_c$.

\section{Mapping the phase diagram in the T, complex $\mu$ plane to the
$T,\mu^2$ plane}
The Gran Canonical Partition function 
$\cal Z(\mu) $ is an even function of $\mu$,  which is 
{real valued} for either real and purely imaginary $\mu$,
and complex otherwise. 

We can  map the complex $\mu$ plane onto the complex
$\mu^2$ plane, and consider ${\cal Z}(\mu^2)$. Note that because of the
symmetries of the partition function this can be done without
any loss of generality. 
  
Then, $\cal {Z}$ is real valued on the real $\mu^2$ axis, complex
elsewhere : the situation is  analogous to  e.g. the
partition function as a function of a magnetic field, which becomes
complex as soon as the external field becomes complex, and the physical
domain (real partition function) is associated with real values
of the couplings.
The critical behaviour of the system is then dictated by the zeroes
of the partition function (Lee-Yang zeros) in the complex $\mu^2$ plane.
The locus of the Lee Yang zeros is thought to be associated
to a general surface of phase separation \cite{janke}, and
phase transition points, for each value of the temperature, 
are associated with the Lee-Yang edge building up in the
infinite volume limit, thus defining  a curve in the $T, \mu^2$ plane.

This simple reasoning shows that it is very natural to re-think
the phase diagram in the T, $\mu^2$ plane, allowing $\mu^2$ to take 
both negative and positive values (Fig. \ref{fig:pd}). The 
critical line itself should be a smooth function
$T(\mu^2)$, making it natural the analytic
continuation from positive to negative
$\mu^2$ values. The vertical dash line is the Roberge-Weiss discontinuity,
ending at $T_{RW}$, 
and it is still not completely clear how it morphs with the chiral transition
\cite{thanks}.

Experience with statistical models shows that not only
the critical line,  but also the critical exponents 
are smooth functions of the couplings \cite{potts} (aside of course
from endpoints, bifurcation points, etc.). Hence,  they can
be safely expanded, either via Taylor expansion or
a suitable ansatz. In particular,
$\mu^2_c=0$ has no special character: it
is just the point where the Lee Yang edge hits the real axis where
$T=T_c$. 

All in all, our task is to simulate the theory in the 
$\mu^2 \le 0 $ strip, and to do our best to 'extrapolate' the results 
to real values of $\mu$. There are two main possibilities : one is via a 
canonical approach, and the other via an analytic continuation.

\section{The Canonical Approach}

An imaginary chemical potential $\mu$ in a sense bridges Canonical and
Grand Canonical ensemble:
\begin{equation}
{\cal Z_C(N)} = \frac {\beta} {2 \pi} \int_0^{2 \pi/ \beta} d \mu {\cal Z_{GC}}
(i \nu) e^{- i \beta \mu N}
\end{equation}
hence,  $Z(V,T,i \mu_I)$ can also be used to reconstruct
the canonical partition function $Z(V,T,n)$ at fixed quark number $n$~\cite{Roberge},
\ie at fixed density:
\beq
Z(V,T,n) &=& {\rm Tr} \left( ( e^{-\frac{H_{\rm QCD}}{T}} \delta(N - n) \right) = 
\frac{1}{2\pi} {\rm Tr} \left( e^{-\frac{H_{\rm QCD}}{T}} \int_0^{2\pi} 
{\rm d} \theta e^{i \theta (N - n)} \right) \nonumber \\
&=& \frac{1}{2\pi} \int_0^{2\pi}
{\rm d} \theta e^{- i \theta n} Z(V,T,i \theta T) \; .
\label{canonical}
\eeq
As $n$ grows, the factor
$e^{- i \theta n}$ oscillates more and more rapidly and the error in the 
numerical integration grows exponentially with $n$: this makes the application
of the method difficult especially at low temperatures where $Z(V,T,i \mu_I)$ 
depends very weakly on $\mu_I$.
The method has been applied in QCD ~\cite{HasTou92} and in
the 2--d Hubbard model~\cite{Dag}~\cite{alford},
where $Z(V,T,n)$ has been reconstructed up to $n = 6$
~\cite{alford}, and more recently again to QCD (see later for some
comparison with other approaches) 
\cite{Kratochvila:2005mk,deForcrand:2006ec,Alexandru:2004dx}
\begin{wrapfigure}{r}{8 truecm}
\centerline{
\includegraphics[width= 8truecm]{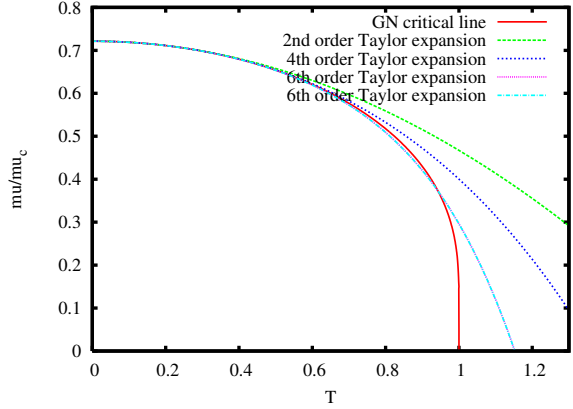}
} 
\caption{The Gross Neveu critical line and its Taylor approximants:
clearly one would need an infinite number of terms to reproduce
the infinite slope of the critical line at $T=0.$ Pade' guarantees a faster
converges.}
\label{fig:Gross_Neveu_simple}
\end{wrapfigure}
\section{Analytic continuation: Taylor, Fourier, Pade'}
In principle, 
if one were able to determine 
thermodynamic observables as a function $F(\mu)$ 
of the imaginary part of $\mu$ with
infinite accuracy, standard complex analysis arguments would 
guarantee that the result will be valid within the 
entire analytic domain. 
Since $F(\mu)$ is not known a priori, 
analytic continuation must rely on some series representation,
or phenomenological insight. 
In the following of this section, 
I will discuss  Taylor series, Fourier series 
and Pade' approximants, which are ratios of polynomials defined as
\begin {equation}
P [N,M] (\mu_i)  = \frac {a_0 + a_1 \mu_i + .....a_M \mu_i^M}{b_0 + b_1 \mu_i
      ..+ b_N \mu_i ^N}
\end{equation}
Note that $P[0,N] =$ is the Taylor N-th partial sum.)

Taylor series was the first idea which came to mind:
one fits the numerical results  to a polynomial, whose coefficients can be
interpreted in terms of a Taylor series centred at $\mu=0$. This allows
an easy contact with the calculations of 
\cite{Gottlieb:1988cq,Bernard:2002yd,Gavai:2003mf,Choe:2002mt,Gavai:2003nn,
Allton:2002zi,Allton:2003vx}
where the various coefficients
are computed as derivatives of the various observables at$\mu=0$.
A truncated polynomial
is of course an analytic function, which can be evaluated everywhere on
the complex plane as a function of the complex variable say $z$:
\begin{equation}
O(\mu) = \sum_k a_k \mu^k 
\end{equation} 
However, convergence will only be achieved within a circle 
(the circle of convergence
of the Taylor series). One can make a virtue of this limitation by
estimating the position of singularities in the complex plane from
the value of the radius of convergence estimated from the behaviour
of the series itself \cite{Gavai:2004sd}. 

We can now ask ourselves, how can we analytically continue beyond the radius
of convergence of the Taylor series. This of course must be possible,
because of the general argument recalled above.

let us remind ourselves that an analytic function is locally
      representable as  a Taylor series. The convergence  disks can be
      chosen is such a way that they overlap two by two, and cover the
      analytic   domain.  Thus,   one  way   to  build   the  analytic
      continuation   is  by  connecting   all  of   these  convergence
      disks. The arcs of the  convergence circles which are within the
      region where f is analytic have a pure geometric meaning, and by
      no means are an obstacle to the analytic continuation.
      Assume now that the  circle of convergence about $z$  = (0,0) has
      radius  unit, i.e.  is  tangent  to the  lines  which limit  the
      analytic domain; take now a $z$ value, say $z_1 = (0,a), 1/2 < a < 1  $
      inside the convergence  disk as  the origin  of a  new series
      expansion, which is explicitly defined by the rearrangement 
                  $(z- z_0)^n = (z - z_1 + z_1 - z_0)^n$ 
      As the  radius of  convergence of the  new series will  be again
      one, this procedure will extend  the domain of definition of our
      original  function (the  two series  define restrictions  of the
      same function to the intersection between the two disks), and by
      'sliding'  the convergence disk  we can  cover all  the analytic
      strip.

An alternative parametrisation, advocated in 
\cite{D'Elia:2002gd,Kratochvila:2006jx,D'Elia:2005qu}  for dealing with the
data in hadronic phase, uses  the Fourier Analysis:
\begin{equation}
O(\mu_I) = \sum_k a_k exp (k i \mu_I) O(\mu) = \sum_k a_k exp (k  \mu)
\end{equation}.

The latest proposal is to use  Pade' 
Approximants \cite{Lombardo:2005ks,Papa:2006jv} . Pade' approximants 
have a time honored history in statistical mechanics,
 and we have borrowed these idea to try and improve the analytic
continuation in our context.
We have sketched above the standard theoretical argument
to demonstrate the feasibility of analytic continuation beyond the
radius of convergence of the Taylor series, and we will show that 
the Pade' series is one practical way to accomplish it.

\section{Interlude : Insight from models}

We collect here a miscellaneous (and by no means
systematic or complete) set of results for models, which might be useful 
to keep in mind for further experimentation.

\subsection{The Gross-Neveu model}

The Gross-Neveu model in three dimensions  is
interacting, renormalizable and can be chosen with
the same global symmetries as those of QCD
which, when spontaneously broken at strong coupling, produce
Goldstone particles and dynamical mass generation. 
As such, the Gross-Neveu model (as well as other four fermion
models) can provide some  guidance to the understanding
of the QCD critical behaviour 
(see e.g.refs. \cite{Hands:2001jn,Hands:1998jg,kli}).

The critical line for the three dimensional GN model
was calculated in ref.~\cite{Hands:1992ck} and reads
\begin{equation}
1 - \mu /\Sigma_0 = 2 T /\Sigma_0 \ln (1 + e^{-\mu/T})
\label{eq:eosgn}
\end{equation}
where $\Sigma_0$ is the order parameter in the normal
phase. Setting $\mu=0$ in the above equation gives
the critical temperature at zero chemical potential, 
$T_c(\mu=0) = \Sigma_0/2 \ln 2 \simeq .72 \Sigma_0$

\begin{wrapfigure}{l}{8 truecm}
\centerline{
\epsfig{file=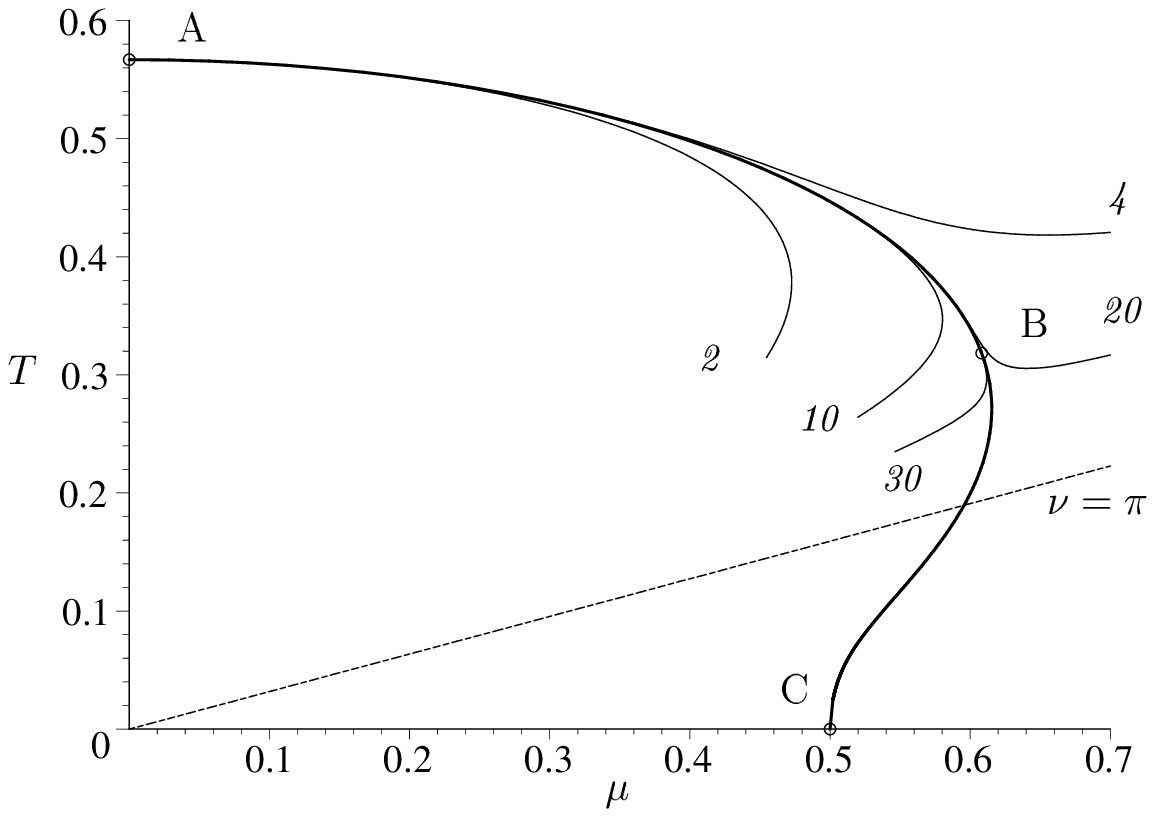,width=5cm,angle=270}}
\caption{2nd order critical line in comparison to 
analytic continuation from imaginary to real $\mu$ 
via power series expansion around $\theta=0$. 
The curves correspond to 2nd, 4th, 10th, 20th, 30th order
in $\theta$ and reach beyond the tricritical point B. 
The series converges above the line $\nu=\pi$.From 
ref.\cite{Karbstein:2006er} }
\end{wrapfigure}
Expanding now $\ln (1+e^{(-x)}) \simeq \ln 2 - 1/2 x + 1/ 8 x^2$,
and eliminating $\Sigma_0$ in favour of $T_c$, we get
\begin{equation}
(T - 1/2 T_c)^2 + \mu^2/(8 \ln 2) = T_c^2 / 4
\end{equation}
This expression can be cross checked with different orders of the Taylor
expansion: it is interesting to mention that, even barring problems
connected with the radius of convergence we cannot expect that
the Taylor expansion describes well the critical line:
note the infinite slope at T=0, and 
we believe the Fig. 2  is self explanatory. 

A more sophisticated study has been carried out recently,
\cite {Karbstein:2006er}, addressing the very interesting question 
of the behaviour of the analytic continuation past a tricritical point,
see Fig. 3: they find that the tricritical point does not limit
the radius of convergence of the critical line, and propose a strategy
based on the analysis of the effective potential for identifying it.

\subsection{Random Matrix Theories}

As it is well known (see e.g. \cite{RMT1}), 
there is a remarkable relation
between the symmetry breaking classes of QCD and the
classification of chiral Random matrix Ensembles.

For QCD with fermions in the complex representation (i.e. $N_c > 2$,
fundamental fermions) with pattern of SSB 
$SU(N_f)_R \times SU(N_f)_L \rightarrow SU(N_f)$, the corresponding
RMT is chiral unitary, $\beta =2$ in the Dyson representation.
On the lattice,  staggered fermions have unusual patterns of 
$\chi SB$: all real and pseudoreal representations
are swapped. However, for complex representations, the corresponding
RMT ensemble remains chiral unitary \cite{RMT1}.
The critical line in the $T, \mu$ plane for this ensemble
derived in ref. \cite{RMT}:
\begin{equation}
(\mu^2 + T^2)^2 + \mu^2 - T^2 = 0
\label{eq:eosrmt}
\end{equation}
is thus valid both on the lattice and in the continuum. 
Expanding it to $O(\mu^2$) we obtain
\begin{equation}
T^2 = T_c^2 - 3 \mu^2  
\end{equation}
and by comparison with the exact result~\ref{eq:eosrmt}, 
we note that this simple expression
describes well the critical line basically till its
endpoint. Again, note that the parametrisation is a Jordan
curve with infinite slope at $T=0$.
\begin{wrapfigure}{l}{8 truecm}
\centerline{
\includegraphics[width= 8truecm]{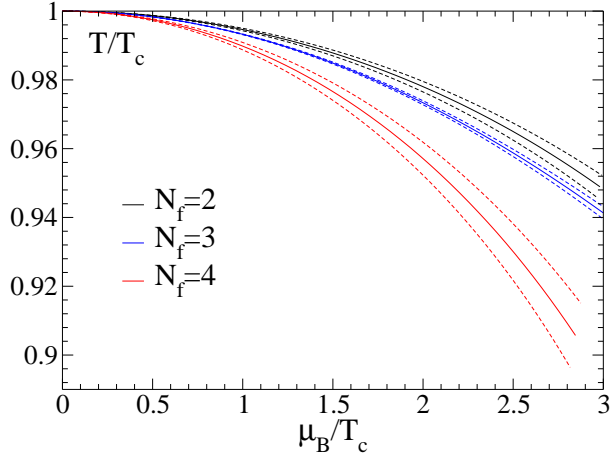}
} 
\caption{
Summary plot\cite{deForcrand:2003hx}  for the critical line for
$N_f=2$\cite{deForcrand:2002ci}, 
$N_f = 3$ \cite{deForcrand:2003hx}, 
$N_f = 4$ \cite{D'Elia:2002gd} from imaginary chemical potential
calculations. }
\label{fig:summary}
\end{wrapfigure}

\subsection{One Dimensional QCD}

Another amusing example is one dimensional QCD , whose free energy
as a function of fugacity can
be analytically computed \cite{Bilic:1988rw}.
In terms of the fugacity $f = e^{3 \mu /T}$ ${\cal Z}$ reads:
\begin{equation}
{\cal Z} (f) = f^2 + 1 +f \cosh (3 m /T) = P[2,2](f)
\end{equation}
We note that the Pade'expansion is exact, while a Taylor series, again ,
will not converge outside its radius of convergence. 
The model is also an excellent testbed for the general pattern of
complex zeros of the partition function: the analysis \cite{talkbnl}
 of such zeros
does indeed confirm the general discussion of \cite{Stephanov:2006dn}.

\section{The critical line at small $\mu$ from the Taylor expansion}
The first studies  of the critical line have indeed found that a simple
polynomial approximation suffices to describe the data, within
the current precision.  The precision is demonstrated
in Fig. 4 \cite{deForcrand:2003hx} 
where the  results on the three flavor model \cite{deForcrand:2003hx}
are superimposed to the  ones 
with two\cite{deForcrand:2002ci} and four flavors\cite{D'Elia:2002gd}.

\begin{figure}[t]
\centerline{
\includegraphics[width= 8  truecm,angle=270]{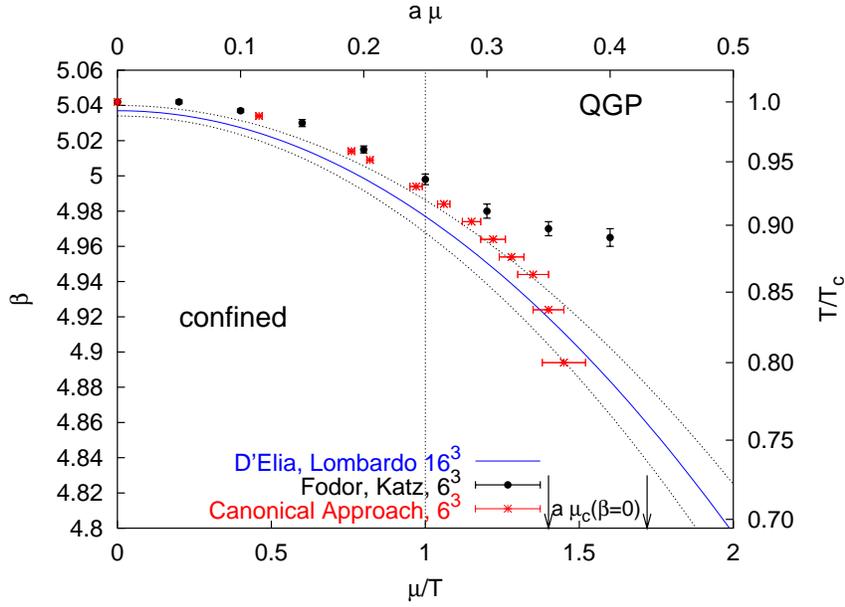}}
\caption{Comparison among different techniques for  the critical line
of four flavor QCD I : 
The analytic continuation relying on the Taylor
expansion cannot be automatically trusted beyond the vertical line.
From Ref. \cite{deForcrand:2006ec}}
\end{figure}

\begin{figure}[bt]
\includegraphics[width=5.9 truecm, angle = 90]{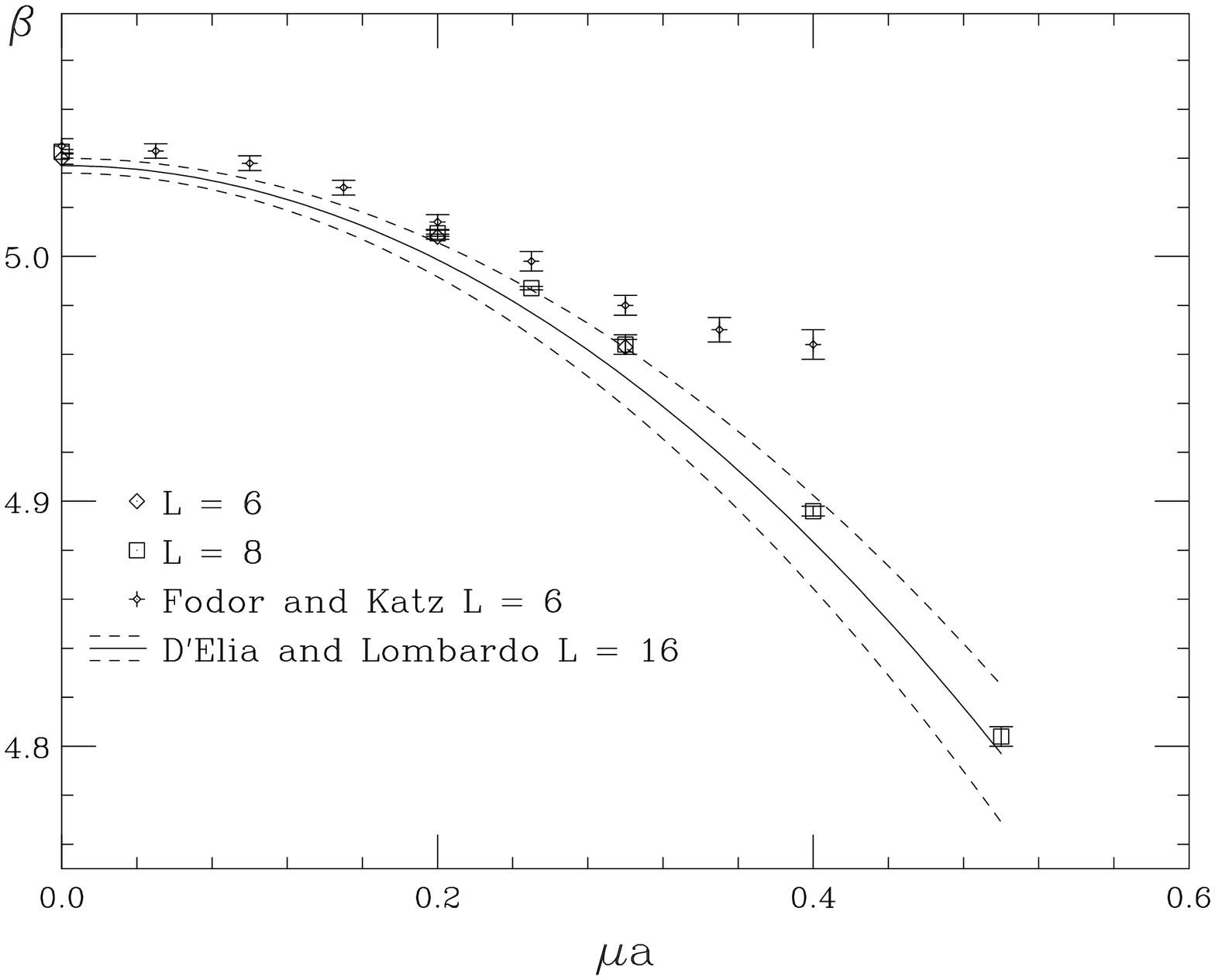}
\includegraphics[width = 8.1 truecm]{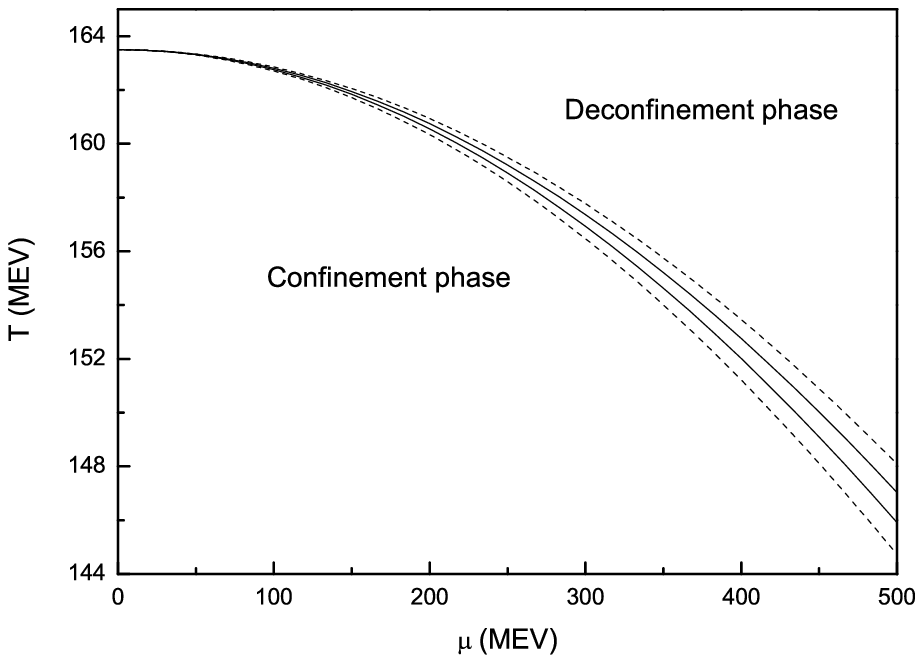}
\caption{Results for the critical line from the generalised imaginary
chemical potential approach of ref. \cite{Azcoiti:2005tv} compared
with those of ref. \cite{D'Elia:2002gd} 
(left); Comparison between Wilson (solid) and staggered (dash) fermions,
from ref. \cite{Chen:2004tb}}
\end{figure}
The overall trend is consistent with the transition becoming weaker with
increasing the number of flavor.

Another nice consistence check can be done by exploiting the four flavor
model, which was
studied in a greater detail, and it is shown in Figs. 5,6.

In Fig. 5 we see a compilation of 
results from different methods. In Fig. 6, left 
the results are contrasted with a generalised imaginary chemical
potential approach \cite{Azcoiti:2005tv}. They define
\begin{equation}
S = S_\mathrm{PG} + ma \sum_n \bar\psi_n\psi_n
+ \frac{1}{2}\sum_n\sum^3_{i=1}\bar\psi_n \eta_i (n)
\left( U_{n,i} \psi_{n+i} - U^\dagger_{n-i,i}\psi_{n-i}\right) +
S_\tau(x, y)\, ,
\label{gaction}
\end{equation}
with
\begin{eqnarray}
S_{\tau}(x, y) &=& x \frac{1}{2} \sum_{n}\bar\psi_n \eta_0 (n)\left( U_{n,0}
\psi_{n+0} - U^\dagger_{n-0,0}\psi_{n-0}\right) \nonumber \\
&+& y \frac{1}{2} \sum_{n} \bar\psi_n \eta_0 (n)\left(  U_{n,0}
\psi_{n+0} + U^\dagger_{n-0,0}\psi_{n-0}\right)\, ,
\label{actiontxy}
\end{eqnarray}
where $x$ and $y$ are two independent parameters. The QCD action is recovered
by setting $x= \cosh(\mu a)$ and $y= \sinh(\mu a)$. By moving in the $x,y$
space it is possible to include low temperature regions within the radius
of convergence of the Taylor series. 
In Fig. 6, right we see the comparison between the results for Wilson fermions
and those for staggered fermions \cite{Chen:2004tb}.
Recently, an interesting study with two flavor of Wilson fermions has
appeared as well \cite{Wu:2006su}. 
Results from different fermion discretions are
indeed very important to control the continuum limit, and
these cross checks nicely demonstrate the robustness of the results

However, the critical line 
 estimated by use of a second order Taylor series,
in principle cannot be trusted beyond the radius of convergence
of the series itself.  We will discuss in the next Section how to improve
the situation  by use of Pade' approximants.

\subsection{The Endpoint}
A crucial issue remains the determination of the
endpoint,  expected of a theory with two plus one flavor.
The first estimate was given within the
reweithgting method 
$T_E= 160 \pm 3.5 MeV$, $\mu_E = 725 \pm 35 MeV$ \cite{Fodor:2001pe}.
Results obtained at imaginary chemical potential and 
with improved precisions did show a subtle dependence
on the mass values, and on the parameters of the algorithm
\cite{Kogut:2006jg,deForcrand:2006pv}.

Such technical limitations are not specific of imaginary chemical 
potential.  This makes mandatory an extrapolation to physical
values of the quark masses, which , in turn, implies a good control
on the continuum limit.

Recently, Forcrand and Philipsen took an alternate approach to the
search of the endpoint : they observed \cite{deForcrand:2006hh} 
that the existence of the
endpoint depends on general features of the critical surface in
the temperature chemical potential and mass plane (Fig.7) . 
They arrived at the conclusion that, for the lattice spacing considered here, 
the curvature of the critical surface is negative, which seems hardly
consistent with the presence of an endpoint.

This behaviour,  if confirmed, 
(and, indeed, the results of 
ref. \cite{Sinclair:2006zm} support these findings)
would be a strong indication that the behaviour
of QCD is indeed much more subtle than that of the four fermion models
with the same global symmetries, which were first used to
suggest the existence of such endpoint in
the $T, \mu$ plane \cite {MishaLattice}. 
Perhaps not surprisingly, given that the
mechanism of chiral symmetry breaking in QCD is associated
with long distance forces, as opposed to the strong local dynamics
which is responsible for chiral breaking in four fermions models.

\begin{figure}[t]
\vspace*{-0.5cm}
\hspace*{-1.5cm}
\includegraphics*[width=0.6\textwidth]{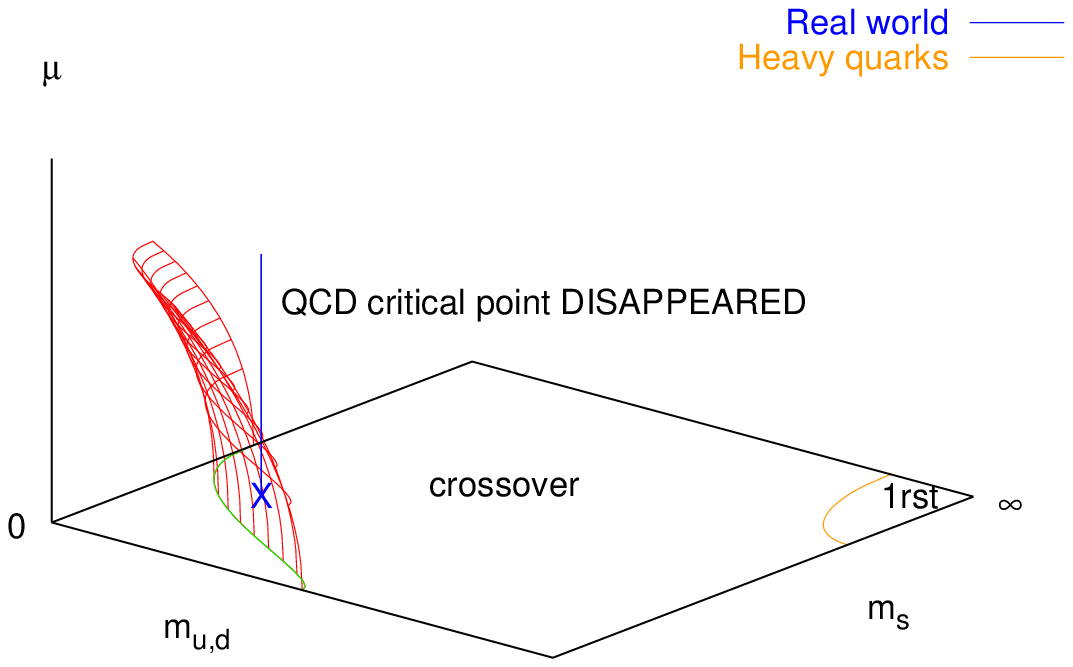}\hspace*{0.25cm}
\includegraphics*[width=0.5\textwidth]{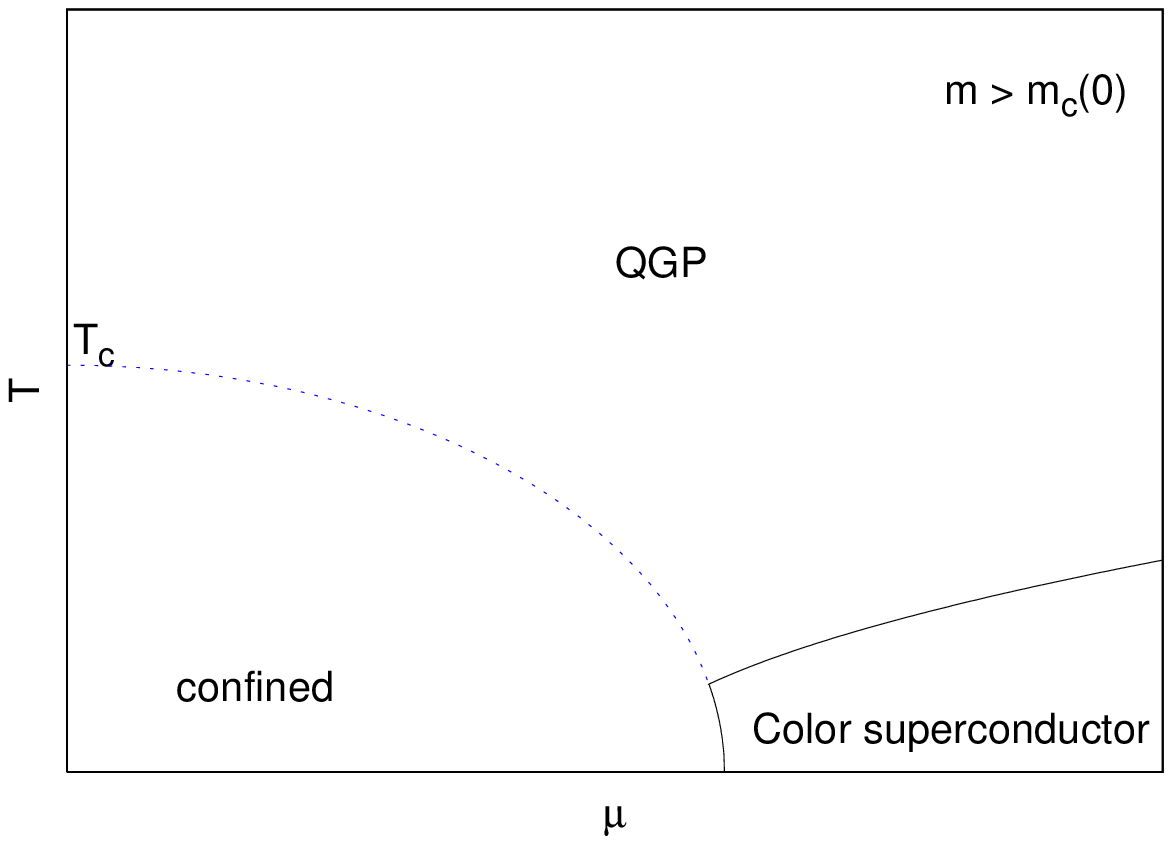}
\caption[]{\em
For $dm_c(\mu)/d\mu^2 <0$, there is no critical point at all, the dotted line on the right is merely a crossover., from Ref. \cite {deForcrand:2006hh}}
\label{nons}
\end{figure}

\subsection{Chiral Symmetry, confinement, topology}
Theoretical issues related with the nature of the critical line can be studied
directly at imaginary chemical potential, without any continuation.

In Ref. \cite{D'Elia:2002gd,D'Elia:2004at}
we have demonstrated the correlation between chiral condensate and
of the Polyakov loop, and argued that
this correlation should be continued at real baryon density:
to this end, we note that 
if $\beta_c(i \mu_I) = \beta_d(i \mu_I) $ over a finite
imaginary chemical potential interval, then the function 
$\Delta {\beta} (i \mu_I) = 
\beta_c(i \mu_I) - \beta_d(i \mu_I)$ is simply continued to be zero over
the entire analyticity domain, thus demonstrating the correlation
between the chiral and the deconfinement transitions ($\beta_c = \beta_d$)
also for real values of $\mu$. We refer to e.g refs. 
\cite{Mocsy:2003qw,Ratti:2006wg} for
an effective Lagrangian discussion of this issue.

It is interesting to briefly mention results in the two colour model:
in ref  \cite{Alles:2006ea} it is shown that Polyakov Loop, chiral condensate
are correlated in the high temperature region
of the phase diagram in two colour QCD, as they are in three colour QCD.
In addition to this, also the topological charge (Fig.8) is correlated with
the other observables, as it was at $\mu=0$: 
all in all, a finite density
of baryons does not change the nature of the transition from the
hadron to the quark gluon plasma phase. The simulations were repeated 
for the Pisa order parameter in ref. \cite{D'Elia:2006xb}.
\begin{wrapfigure}{r}{7 truecm}
\vskip .5 truecm
\includegraphics[width= 6.5 truecm]{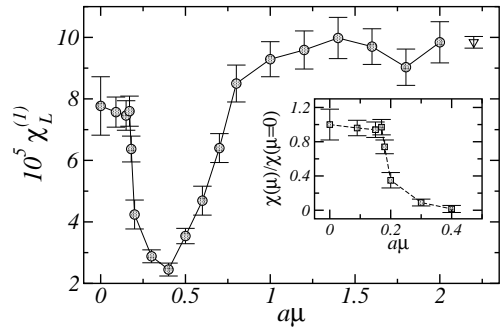}
\caption{Topological susceptibility
 as a function of the chemical potential $\mu$ \cite {Alles:2006ea}.}
\vskip -1 truecm
\end{wrapfigure}

\section{Beyond small $\mu$ via Pade' approximants}
In Fig. 9 we present the Pade' analysis \cite{Lombardo:2005ks} of data
for the critical line of four flavor QCD
(numerical results are from\cite {D'Elia:2002gd} )
Results seem stable beyond $\mu_B = 500 MeV (\mu_B/T \simeq 1)$, 
with the Pade' analysis  in good agreement with Taylor expansion
for smaller $\mu$ values. At larger $\mu$ the Taylor expansion 
seems less stable, while the Pade' still converges, giving a slope 
of the critical line larger than the naive continuation of the
second order Taylor approximations:
we underscore that the possibility of analytically continue the
results beyond the radius of convergence of the Taylor series
by no means imply that one can blindly extrapolate a lower Taylor order
approximation!

 The same bending towards lower $\mu$ values
 is suggested by
recent results within the canonical approach\cite{deForcrand:2006ec} 
and the DOS method \cite{Schmidt:2005ap}, and it agrees with the qualitative
features of the simple models discussed above.

Finally, we summarise in Fig. \ref{fig:foursummary} 
the overall results for the critical line
of four flavor QCD (note that Fig. 9 uses $\mu_B$ while
Fig. 10 uses $\mu$ : the red lines is the same in the two plots).

\subsection{ Thermodynamics beyond $\mu/T \simeq 1$}
Similarly, it is possible to apply the Pade' analysis
to thermodynamics observables.
The Pade' approximants to the results for the chiral
condensate in the hot phase are shown 
in Fig. 11, upper diagram, from \cite {Lombardo:2005ks}. 
We see that the Pade' approximants converge well beyond the would-be radius
of convergence of the Taylor expansion. Similar results have been obtained
in two colour QCD \cite{Papa:2006jv} where it is also possible a direct 
comparison with exact results, in the spirit of ref. \cite{Giudice:2004se}.

We underscore that,  as noted in \cite{D'Elia:2004at} the 
radius of convergence 
should tend to infinite in the infinite
temperature limit, and indeed it has been estimated to be large 
by the Bielefeld--Swansea collaboration\cite{Allton:2005gk} : by
increasing T Pade' and Taylor should become equivalent.

\begin{wrapfigure}{l}{8 truecm}
\centerline{
\includegraphics[width= 7.5truecm]{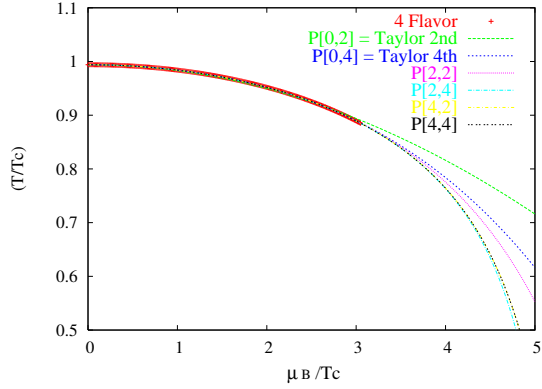}
} 
\caption{Pade' approximants for the critical line of four 
 flavor QCD \cite{Lombardo:2005ks}.}
\end{wrapfigure}

\section{The Hadronic Phase and the Fourier Analysis}

 The grand canonical partition function of the
Hadron Resonance Gas model\cite{Karsch:2003zq,Toublan:2004ks} 
has a simple hyperbolic cosine behaviour. This
can be framed in our discussion of the phase diagram in the 
temperature-imaginary 
chemical potential plane which suggests
to use Fourier analysis in this region, as observables are periodic
and continuous there\cite{D'Elia:2002gd}.
\begin{wrapfigure}{l}{8 truecm}
\vskip .8 truecm
\centerline
{\epsfig{file=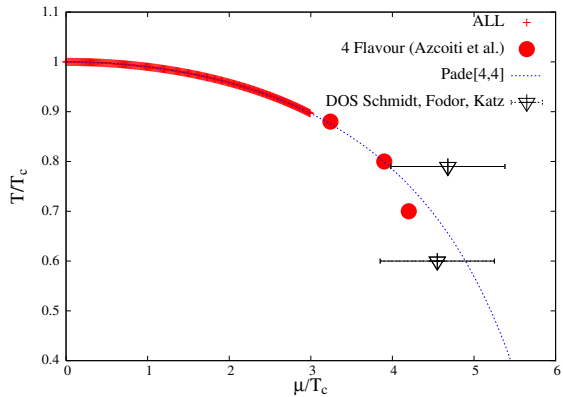 ,width= 7.5 truecm}}
\caption{The critical line of four flavor QCD : the red line is the
small chemical potential region amenable to a Taylor expansion.
Using it as an input to the Pade' analysis, 
the results have been extended deep into
the low temperature phase, in good agreement with other approaches}
\label{fig:foursummary}
\vskip -1.5 truecm
\end{wrapfigure}
For observables which are even ($O_e$) or odd ($O_o$) under
$\mu \to -\mu$    the analytic continuation
to real chemical potential of the Fourier series read
$O_e[o](\mu_I, N_t)   =  \sum_n  a_{F}^{(n)} \cosh [\sinh](
n N_t N_c \mu_I)$.
In our  Fourier analysis of the chiral condensate
\cite{D'Elia:2002gd}
 and of the number density\cite{D'Elia:2004at} - 
even and odd observables, respectively -  
we limited ourselves to $n=0,1,2$ and we assessed the validity
of the fits via both the value of the $\chi^2/{\rm d.o.f.}$ and the stability
of  $a_{F}^{(0)}$ and $a_{F}^{(1)}$ given by one and two cosine 
[sine] fits:
we found that one cosine [sine] fit  describes reasonably well
the data up to $T \simeq 0.985T_c$ (Fig. 12a); 
further terms in the expansion 
did not modify much the value of
the first coefficients and does not particularly  
improve the $\chi^2/{\rm d.o.f.}$:
the data are well approximated by the hadron resonance gas prediction
$\Delta P \propto (\cosh (\mu_B/T) - 1)$
in the broken phase up to $T \simeq 0.985 T_c$.
This behaviour has been confirmed 
by an improved analysis
 in \cite{Kratochvila:2006jx},  Fig. 13.
 The analysis of the
corrections requires better precision \cite{thanksF}.

The analytic continuation (Fig. 12b) of any 
observable $O$ is valid within the analyticity domain, i.e.
till $\mu < \mu_c(T)$, where $\mu_c(T)$ has to be measured independently.
The value of the analytic continuation of $O$ at
$\mu_c$, $O(\mu_c)$, defines its critical
value. When $O$ is an order parameter which is zero in the quark gluon
plasma phase,  the calculation of $O(\mu_c)$
allows the identification of the order of the phase transition:
first, when $O(\mu_c) \ne 0$, second, when $O(\mu_c) = 0$
\cite{D'Elia:2002gd,D'Elia:2004at}. 

\begin{figure}
\centerline{
{\includegraphics [width = 8 truecm]{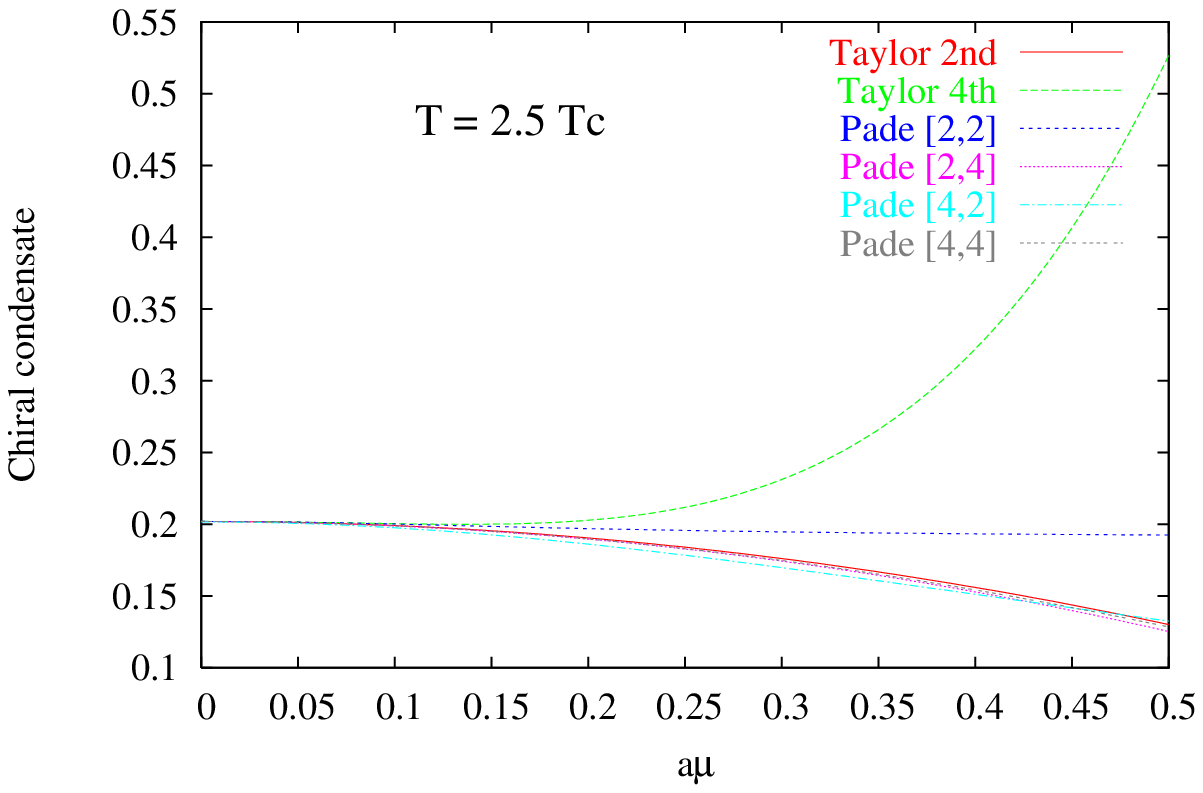}} }
\centerline{
{\includegraphics [width = 6 truecm, angle = 270]{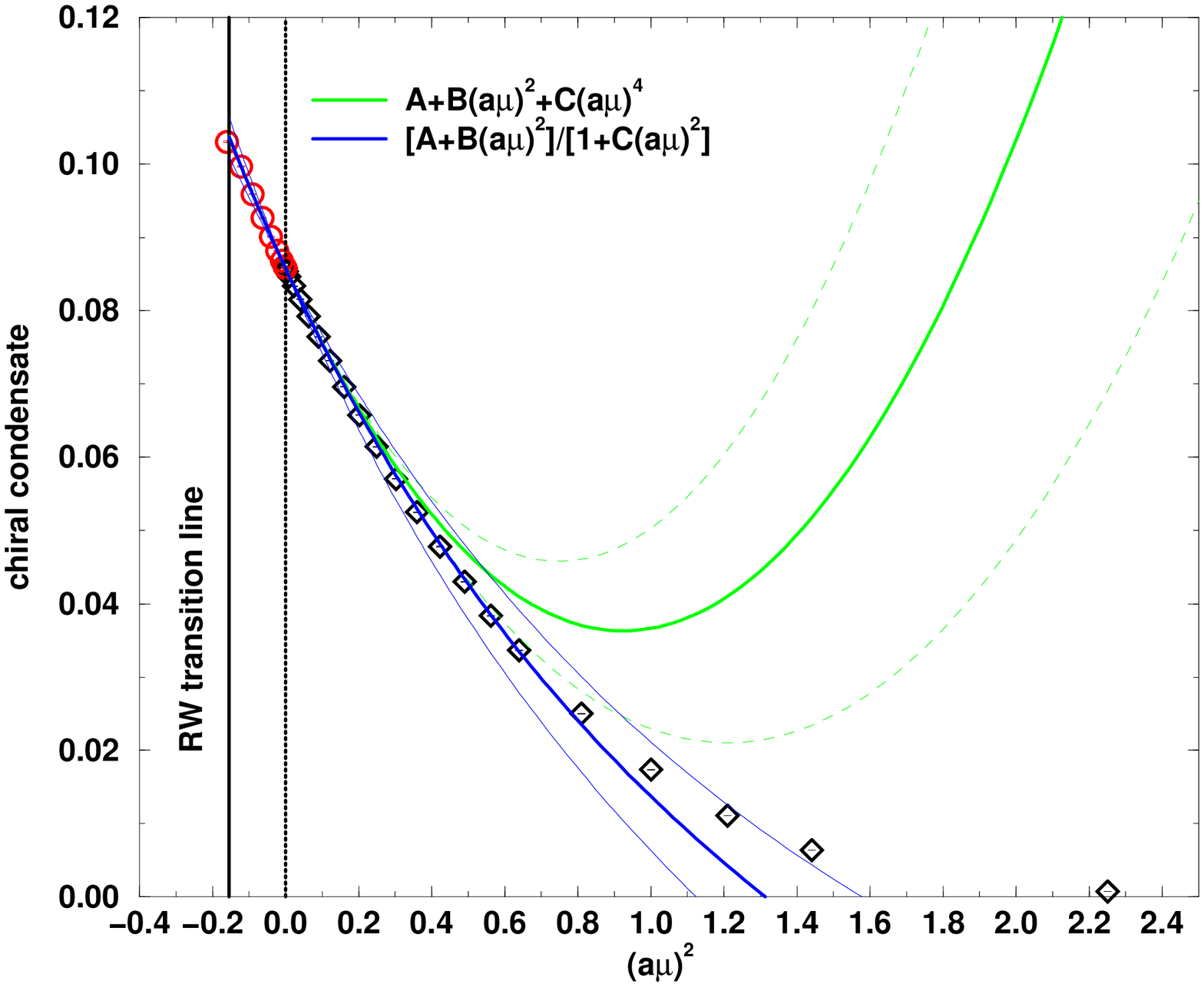}}}
\caption{Analytic continuation via Pade'  approximants for 
the chiral condensate: three colour QCD (top diagram) \cite{Lombardo:2005ks} 
and two colour QCD (bottom diagram) \cite{Papa:2006jv}. For real QCD convergence seems to be achieved for Pade'[N,M], $N+M \ge 6$. For two colour QCD Pade'[2,2]
suffices. }
\end{figure}
\begin{figure}
{\epsfig{file= 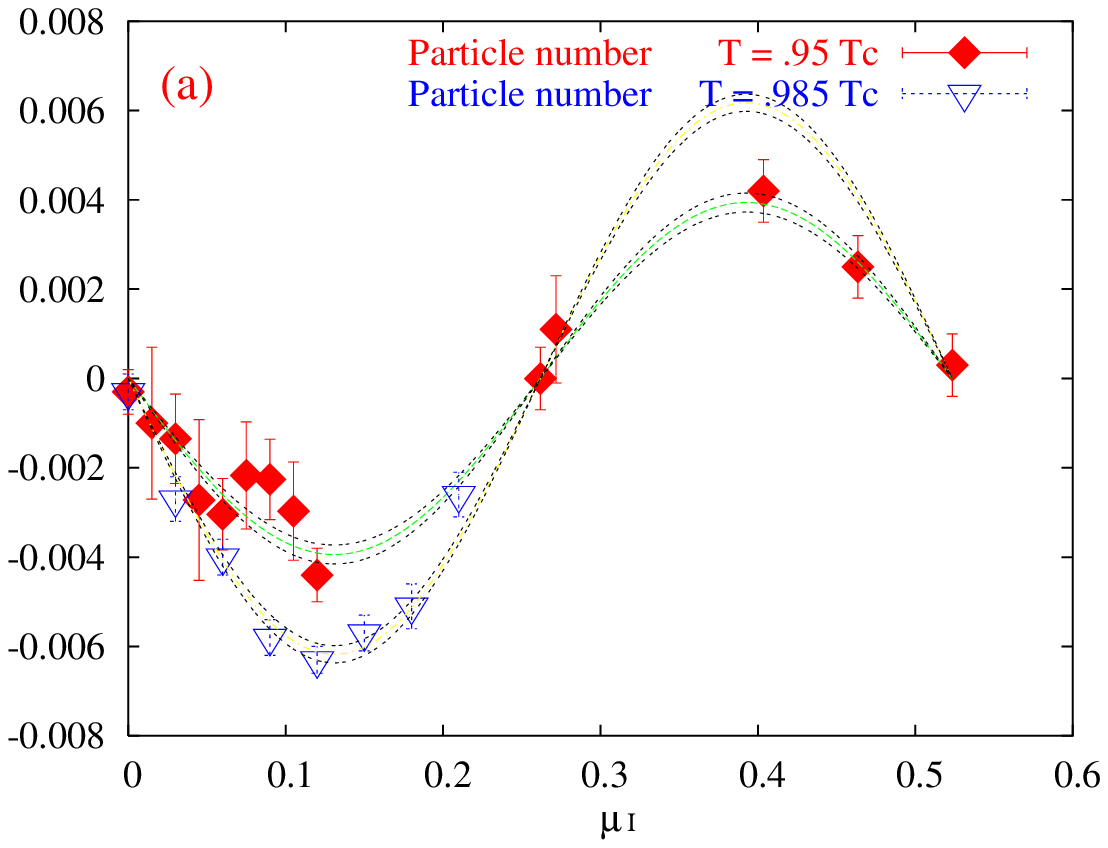, width= 8 truecm}}
{\epsfig{file= 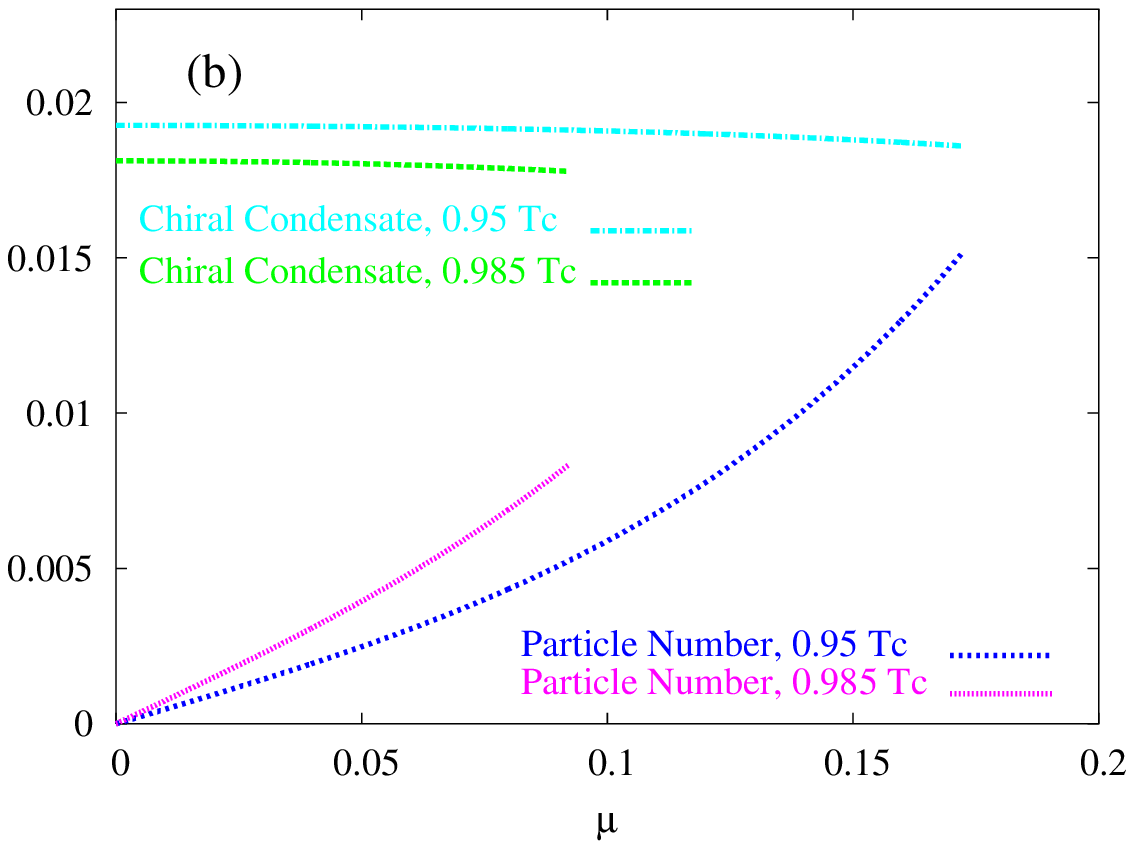, width= 8 truecm }}
\caption{Hadronic Phase:
(a) One Fourier coefficient fit to the particle number, showing
that the Hadron Resonance Model is adequate to describe this data.
(b) Compilation of the results for the chiral
condensate and the particle number as a function
of real chemical potential: the lines are cut in correspondence
with $\mu_c$, showing the first order character
of the phase transition (inferred from the 
chiral condensate) and the critical density \cite{D'Elia:2004wk}.}
\end{figure}

\begin{figure}
\centering
\includegraphics[width=5cm,angle=-90]{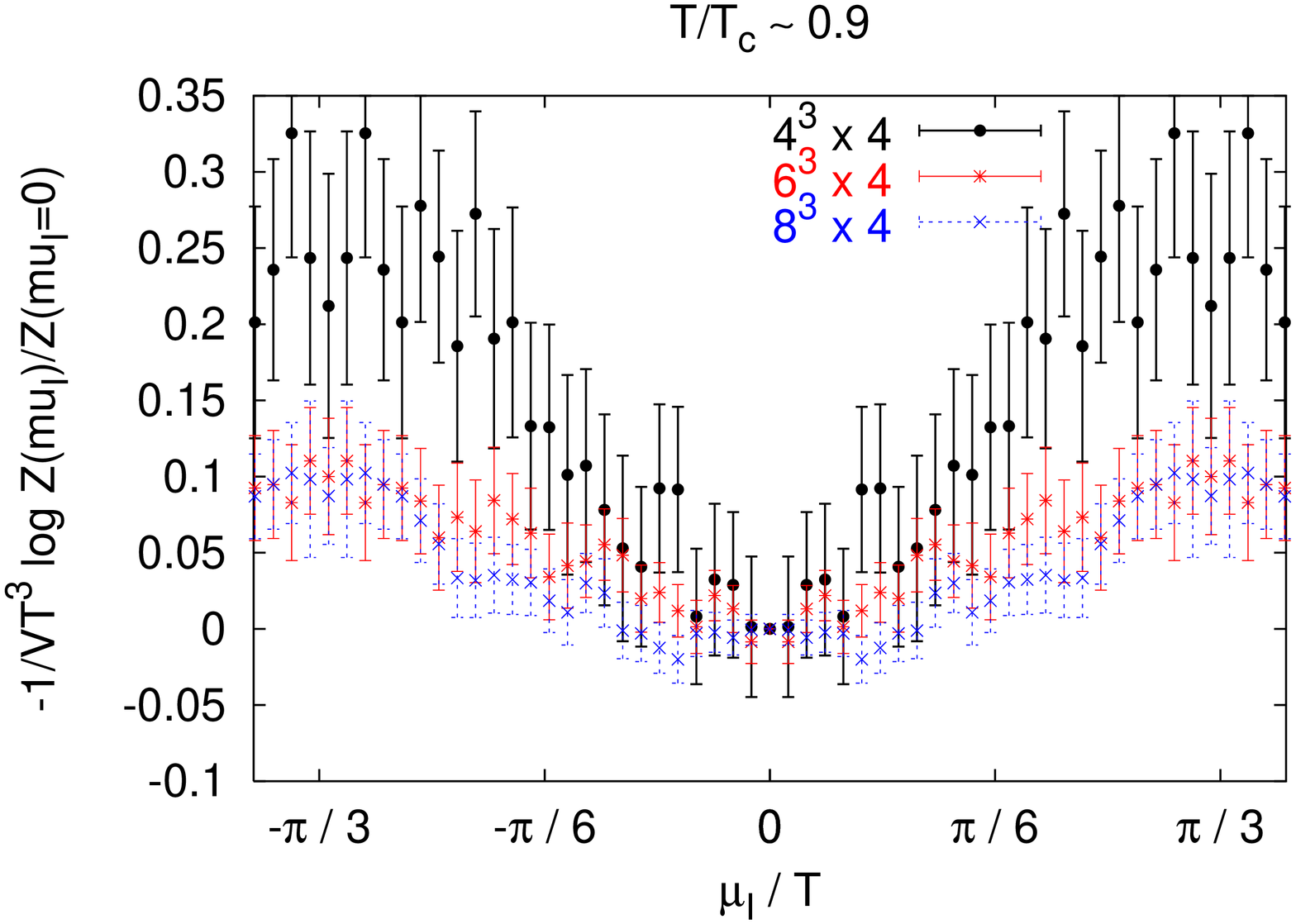}
\includegraphics[width=5cm,angle=-90]{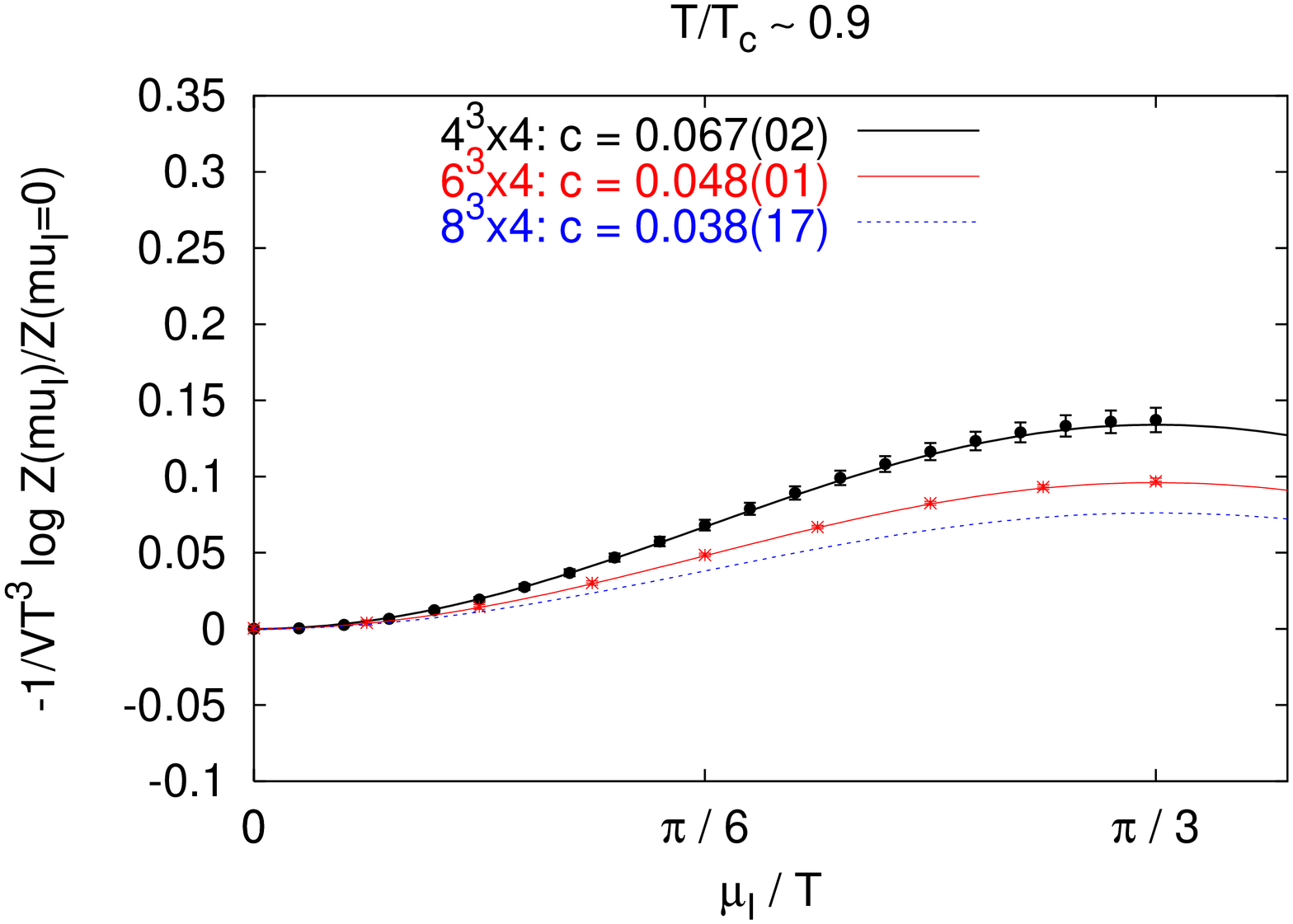}
\caption{$\frac{\Delta F(T,\mu_I)}{V T^4}$ as a function of $\frac{\mu_I}{T}$ for $\frac{T}{T_c} \sim 0.9$. The results from an improved analysis
at imaginary chemical potential confirm that the data
are well accounted for by one component Fourier fit \cite{Kratochvila:2006jx}.}
\vskip .5 truecm
\end{figure}

\section{The Hot Phase and  the QGP}
\begin{wrapfigure}{r}{8.5 truecm}
\centerline{{\epsfig{file= 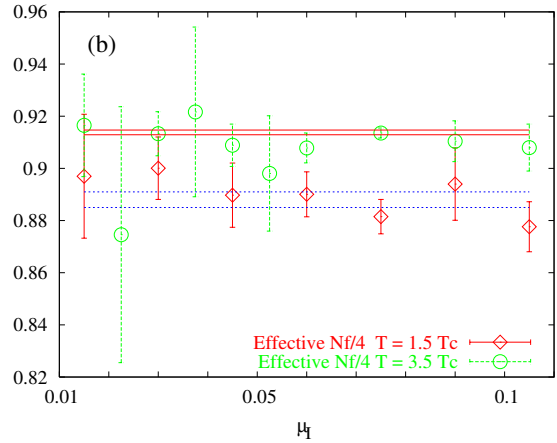, width= 8.5 truecm}}}
\caption{$T \ge 1.5 T_c$: Ratio of the lattice results to the 
lattice free field: the deviation from free field can be 
simply described
by an effective reduced number of flavor, from
ref. \cite{D'Elia:2004wk}}
\end{wrapfigure}
The behaviour of the number density (Fig. 14, from ref.\cite{D'Elia:2004wk})  
approaches the lattice
Stefann-Boltzmann prediction, with some residual deviation.
The deviation  from a free field behaviour can be parametrised as 
\cite{Szabo:2003kg,Letessier:2003uj}
\begin{equation}
\Delta P (T, \mu)  =  f(T, \mu) P^L_{free}(T, \mu) 
\end{equation}
where $P^L_{free}(T, \mu)$ is the lattice free result for the pressure.
For instance, in the discussion of Ref. \cite {Letessier:2003uj}
\begin{equation}
f(T, \mu) = 2(1 - 2 \alpha_s/ \pi)
\end{equation}
and the crucial point was that $\alpha_s$ is $\mu$ dependent.

We can search for such a non trivial prefactor $f(T, \mu)$ by taking 
the ratio between the numerical data and the lattice
free field result $ n^L_{free}(\mu_I)$  at imaginary chemical potential:
\begin{equation} 
R(T, \mu_I) = \frac{ n(T,  \mu_I)}{n^L_{free}( \mu_I)}
\end{equation}
A non-trivial (i.e.
not a constant) $R(T, \mu_I)$ would indicate a non-trivial 
$f(T, \mu)$.
In Fig. 14  we plot $R(T, \mu_I)$ 
versus $\mu_I/T$: the results for $T \ge 1.5 T_c$ seem 
consistent with a free lattice
gas, with an fixed effective number of flavors 
$N^{eff} =   0.92 (0.89) \times 4$ for $T=3.5 (1.5) T_c$ :
in short, the deviation from free field are probably trivial,
and certainly $\mu$ independent.

\newpage
\section{ The non--perturbative Quark Gluon Plasma,  and the critical line of 
QCD in the T,$\mu_I$ plane. }

When temperature is not much
larger  than the critical temperature -- say, $Tc < T < \simeq 2 Tc$ --
strong interactions among the constituents give rise to 
non--perturbative effects: in short, at large T the QGP
is a gas of nearly free quarks, 
which becomes strongly interacting at lower temperatures $T=(1-3)\,T_c$,
see e.g. \cite{Shuryak:2006se,Blaizot:2006up} for recent reviews and
a complete set of references.

Several proposals have been made to characterise the properties of
the system in such non perturbative phase. For instance the above mentioned
strong interactions might be enough to preserve bound states above $T_c$,
while coloured states might appear, deeply
affecting the thermodynamics of the system
\cite{Shuryak:2004tx}. High temperature expansions  are being
refined more and more , so to be able to capture the features of dense systems
down to $T_c$ (see e.g \cite{Vuorinen:2003fs, Ipp:2003yz,Ipp:2006ij}. 
Model theories of quasiparticle physics have been considered
as well\cite{Bluhm:2006av}.

Here we would like to frame the discussion of the strongly interacting QGP
in the context of the critical behaviour at imaginary $\mu$ \cite{appear}.
Consider again  the phase diagram in the T, 
$\mu^2$ plane(Figure 1): we note that
the candidate sQGP region right above Tc is limited by a chiral transition
at negative $\mu^2$ : it is then all  a critical region,
whose features might well carry over to real baryochemical potential.

\begin{figure}[b]
\centering
\includegraphics[width=5cm,angle=-90]{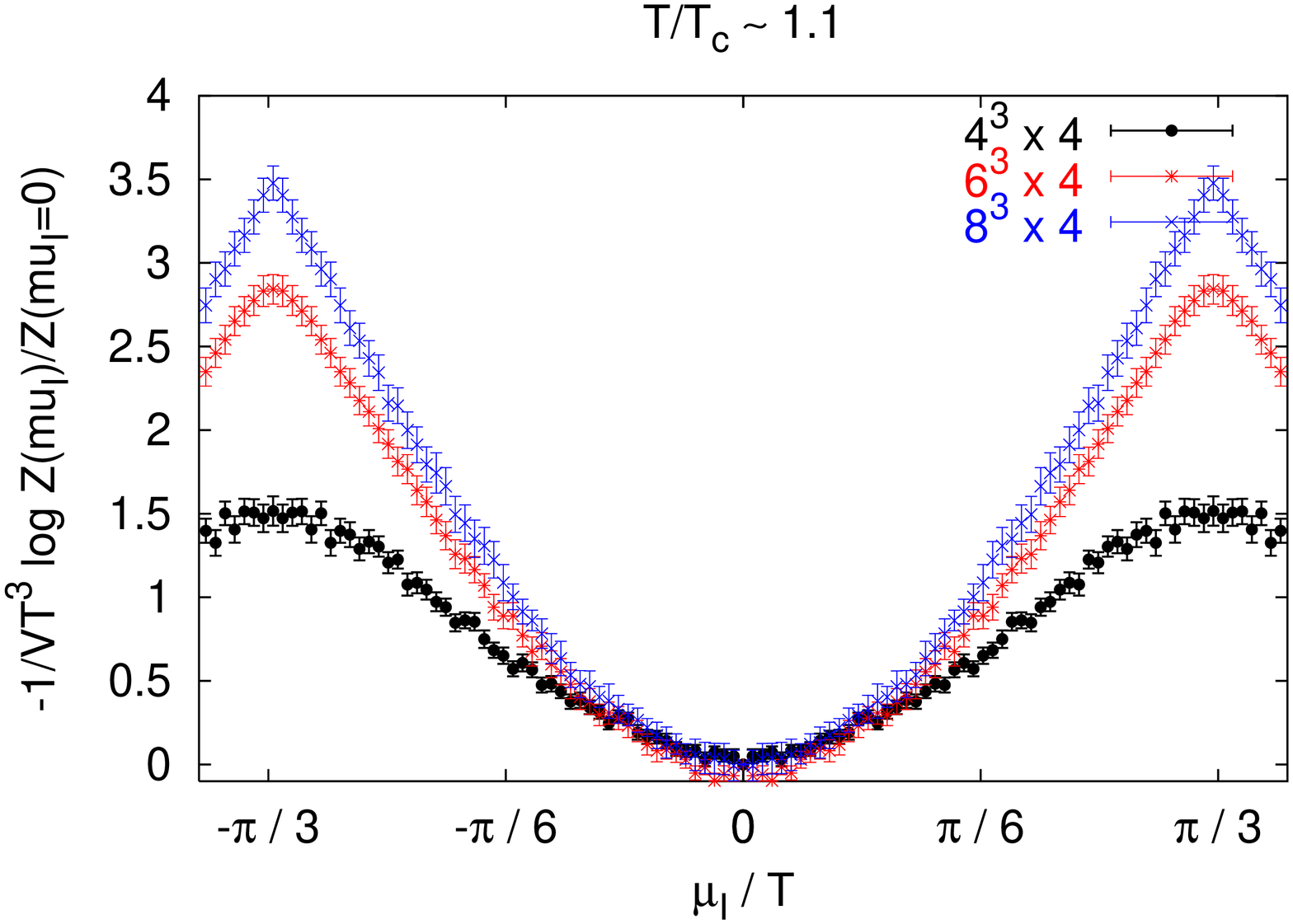}
\includegraphics[width=5cm,angle=-90]{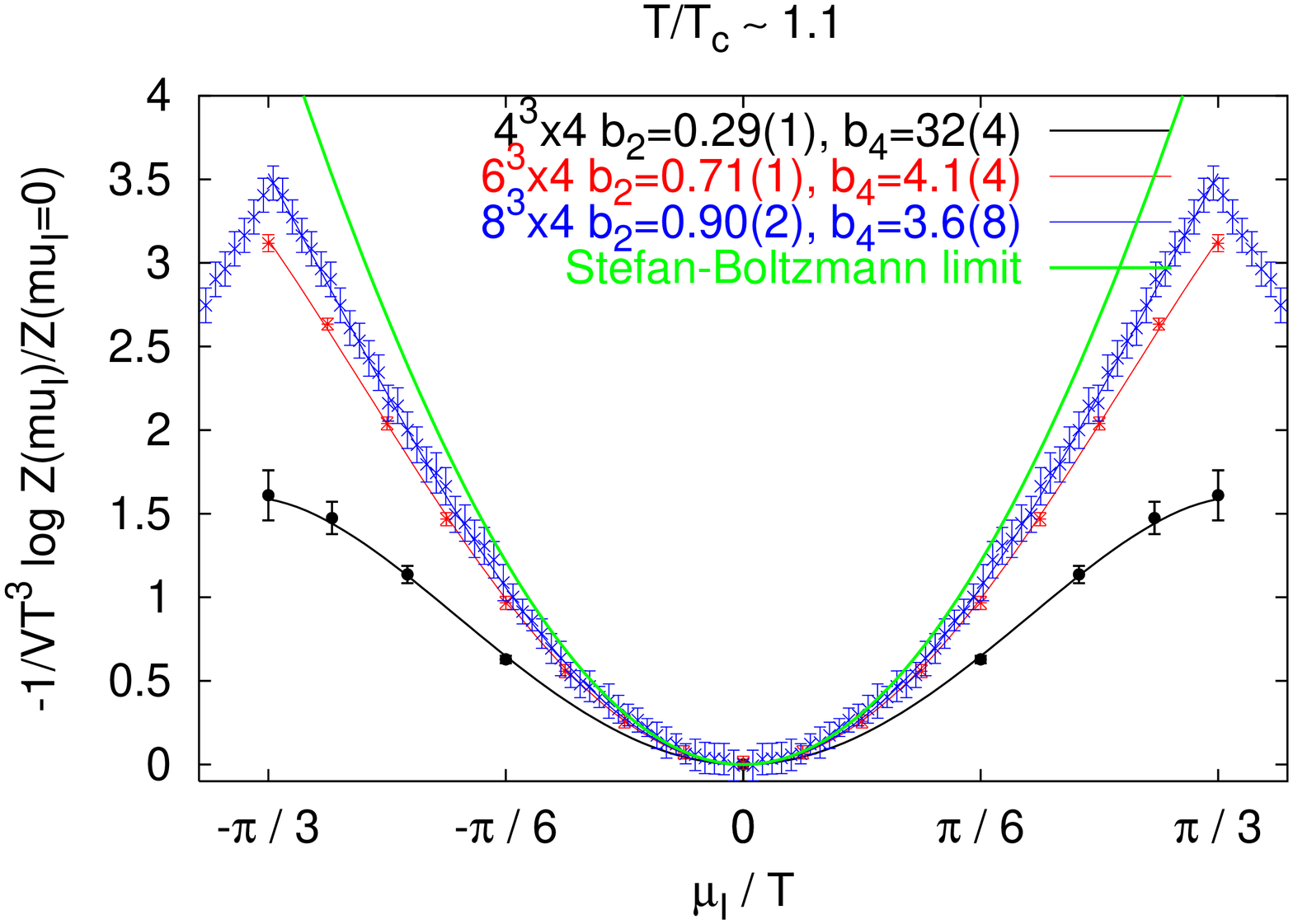}
\caption{$\frac{\Delta F(T,\mu_I)}{V T^4}$ as a function of $\frac{\mu_I}{T}$ for $\frac{T}{T_c} \sim 1.1$. 
({\em left}) The histogram method; ({\em right}) the reweighting method~\cite{Kratochvila:2005mk}, 
supplemented by the histogram results for $8^3 \times 4$. A simple modification of the free gas expression
describes all the data.
As the volume increases, the data come close to the Stefan-Boltzmann limit ($T \to \infty$)
even though $\frac{T}{T_c} \sim 1.1$ only.  Figure and caption are from 
\cite{Kratochvila:2006jx} }
\label{fig:results_chempot_510}
\end{figure}
\subsection{Evidence for strong interactions in the plasma}
In this final subsection I report on our attempt to confront these
speculative idea with lattice data\cite{appear}. 

First, consider the data against a simple free field behaviour:
as done in Fig. 15, we
take the ratio between the numerical results and the free field
results $n(\mu_I)/n(\mu_I)_{free}$(Fig. 16,left)
\cite{D'Elia:2005qu,appear}:
$n(\mu_I)/n(\mu_I)_{free}$ is far from being
constant (as opposed to the findings at higher temperature, Fig. 14), 
which cannot be accounted for by any simple renormalisation 
of the degrees of freedom.

In ref. \cite{Kratochvila:2006jx}, Fig. 15, it was confirmed that
the coefficients of a polynomial fit are not in a simple relations
with those of a lattice free field.
 
\begin{figure}
{\epsfig{file=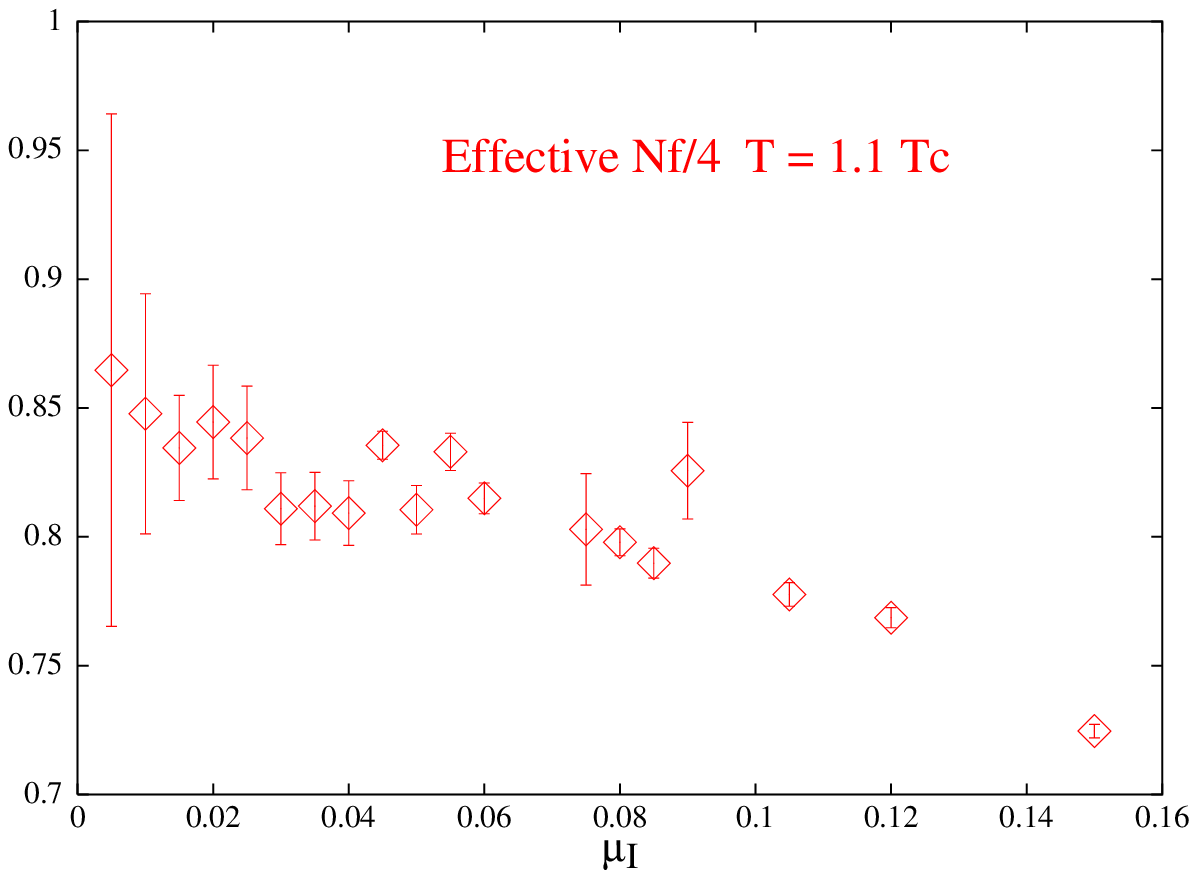,width= 8 truecm}}
\epsfig{file=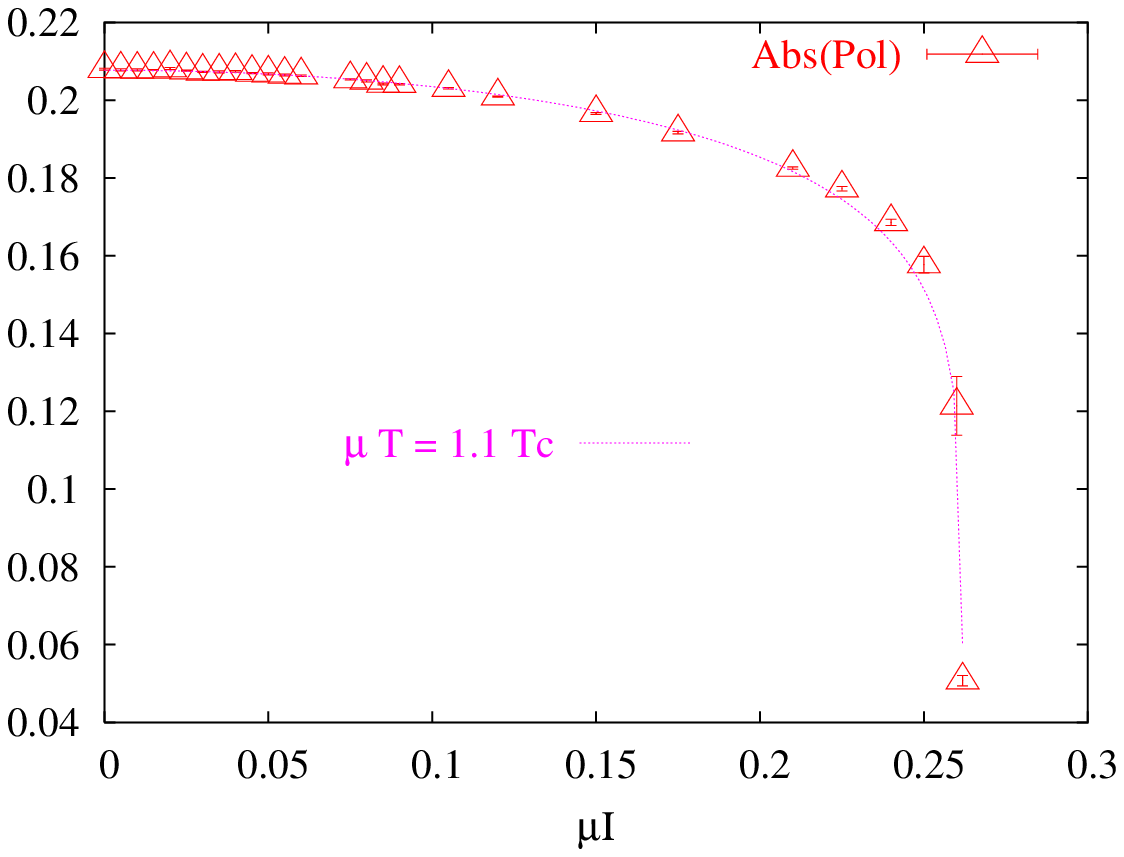,width=8 truecm}
\caption{$T = 1.1 T_c$ $n(\mu_I)/n(\mu_I)_{free}$ as a function of $\mu_I$ ,
showing a very clear evidence of a deviation from a free field behaviour(left diagram). Average Polyakov loop fitted to the form predicted by a simple
critical behaviour  $f_c(\mu_I)$ \cite{D'Elia:2005qu,appear}}
\end{figure}
Next, we checked our data against the form proposed in ref. 
\cite{Ejiri:2005wq}, 
which was motivated by the hadron resonance gas parametrisation.
Going imaginary (and setting $\mu_{isospin}=0$ ) 
the proposal of ref. \cite{Ejiri:2005wq} yields:
\begin{equation}
\frac{P}{T^4} =  A_q(T)\cos(\mu/T)) + B_qq(T) \cos(2 \mu/T) + C_{qqq} (3 \mu /T)\end{equation}
giving in turn:
\begin{equation}
n(i \mu,T) = A_q(T) sin (\mu/T) + 2 B_{qq} (T) sin (2 \mu/T) + 3 C_{qqq} (3 \mu /T)
\end{equation}
We have  performed different fits to the above form, none of them
really satisfactory.   More terms improve of course the fit,
but we do not get any close to the nice agreement with HRG observed in
the hadronic phase, and in particular it is difficult to capture the behaviour
close to the RW singularity.

We propose to use a  power law fit derived from a singular behaviour
of the free energy:
\begin{equation}
log Z \propto \frac {1}{({\mu_I^c}^2 - \mu_I^2)^\alpha} 
\end{equation}
This in turn gives
\begin{equation}
Pol(\mu_I) \propto ({\mu_I^c}^2 - \mu_I^2)^{(\beta)}
\end{equation}.
In Figure 16, right we show the results of the fit, which looks indeed
satisfactory \cite{appear}.

From the results above, we conclude
that the data in the candidate region for a strongly coupled QCD
are very well accounted for by a conventional critical behaviour:
clearly, a free field behaviour would have been incompatible 
with it.
In other words, the nonperturbative features of the plasma are closely
related with the occurrence of the critical line at negative $\mu^2$!

We can then go back and analyse the behaviour of $n(\mu)$: the results
for the Polyakov loop validates the simple form for the free energy,
which, in turn, gives for $N(\mu_I)$ 
 \begin{equation}
 n(\mu_I) \propto \mu_I ({\mu_I^c}^2 - \mu_I^2)^{(\alpha - 1.)}
\end{equation}
and we checked that the behaviour of $n(\mu)$ is reasonably
well reproduced by our fit.

In turn, it is now possible to analytically continue from imaginary to real
chemical potential, yielding:
\begin{equation}
 n(\mu) = K( \mu ({\mu_I^c}^2 + \mu^2)^{(\alpha - 1.)})
\end{equation}
where $\alpha = 1.2$.

It is then amusing to notice that by using the simple 
arguments from the theory of critical phenomena we arrive at a modified
Stefann-Boltzmann law, which would correspond to $ \alpha = 2$.(modulo
coefficients).

Obviously, this modified form accounts for a slower increase of the
particle density closer to $Tc$ than in the free case, as well as
for the behaviour observed in fig.16, left.

From a more mathematical point of view, the proposed parametrisation
is a Pade' approximant of order [2,1], as appropriate in the standard
application to critical phenomena. The results thus obtained can be 
analytically continued within the entire analyticity domain.

We underscore that our results do not rule out the
HRG-type parametrisation proposed in ref. \cite{Ejiri:2005wq}, 
and, in turn,  the 
occurrence of coloured states: the HRG form might still be valid,
once a dependence on the chemical potential of $A(T), B(T), C(T)$ 
is considered. 

On the other hand, if this is the case, the simple interpretation of
susceptibility ratios as probes of degrees of freedom 
has to be revised. 

\section{Summary}

It is possible to study QCD for imaginary values of the baryochemical
potential, and infer  properties of the physical region.
From a technical point of view, these simulations are no more
expensive than ordinary QCD, and, in particular, the infinite volume
limit is well defined. On the other hand, we should be aware that
we are extrapolating results: 
one has to pay attention to the fact that,
by modifying the fitting function (whatever it might be) at imaginary
chemical potential  by a non--leading term,
the difference in physical quantities is still non leading.
This can be complicated, and often mathematical arguments need to be
supplemented by some physical insight.

We have seen that by use of different techniques and
lattice discretization we can gain a reasonable control
on the shape of the critical line of two, three, two plus one, and four
flavor QCD, as well as on thermodynamics observables in different phases. 
The results can be pushed beyond the
radius of convergence of the Taylor expansion, and 
the density of states method or canonical approach, even if more preliminary,  
offer some guidance from the low temperature side. 

A cross check with analytic models is particularly simple, a
one can analytically continue them from real to imaginary chemical
potential.  
Results obtained in this way include a verification of the validity 
of the hadron gas model,  and the approach to a free gas. In the hot 
phase close to $T_c$   the critical line at imaginary
chemical potential suggests an alternative interpretation of the 
non perturbative features of the strongly interactive quark gluon plasma,
which is consistent with the numerical results.

\end{document}